\documentclass[prb]{revtex4}

\usepackage{graphicx}
\usepackage{dcolumn}
\usepackage{bm}

\begin{document}


\title{Measuring of fissile isotopes partial antineutrino spectra in direct experiment at nuclear reactor}

\author{V.V. Sinev}
\affiliation{%
Institute for Nuclear Research RAS, Moscow\\
}%

\date{\today}

\begin{abstract}
The direct measuring method is considered to get nuclear reactor antineutrino spectrum. We suppose to isolate partial spectra of the fissile isotopes by using the method of 
antineutrino spectrum extraction from the inverse beta decay positron spectrum applied at Rovno experiment. This admits to increase the accuracy of partial antineutrino spectra 
forming the total nuclear reactor spectrum. It is important for the analysis of the reactor core fuel composition and could be applied for non-proliferation purposes.
\end{abstract}

\maketitle

\section*{Introduction}

Energy spectrum of antineutrinos from nuclear reactor is a fundamental characteristic of a reactor. When outgoing from a reactor antineutrinos penetrate through 
any shielding. These particles carry out the information concerning the chain reaction in the reactor core. Their spectrum is unique for every reactor type and 
depends on the reactor fuel composition. That is why just when the difference in energy spectra of the fuel components became clear there appeared an idea 
of reactor control through the neutrino emission [1].

There are mainly four fissile isotopes which undergo the fission in the core, $^{235}$U, $^{239}$Pu, $^{238}$U and $^{241}$Pu. Others give an input less than 
1\% and may be neglected. Detector at some distance can detect the total flux from all these components. But during the reactor operational run the shape of total 
spectrum changes because of the burn up effect. So, one can know the fuel composition of the reactor by fitting the total spectrum with the sum of four partial spectra. 
But what are the uncertainties of these partial spectra.

At the beginning of the era of experiments with reactor neutrinos the partial spectra of individual isotopes were calculated [2, 3], but the accuracy of the calculations 
was not so good. Mainly because of insufficient knowledge of fission fragments antineutrino spectra. Later the situation becomes better, while the base of fragments was 
growing [4, 5, 6]. It becomes much better when the first experimental spectra appeared. They were obtained by converting measured beta-spectra from fissile isotopes 
[7, 8, 9]. Electron spectra were measured at Grenoble in ILL by using magnetic spectrometer. These spectra are accounted as the best. Their uncertainty in main 
spectrum part (2$-$7 MeV) is 3.8$-$4.2\% at 90\% CL.

In spite of high enough accuracy this is not sufficient, reactor control demands at least 1
control problem this accuracy needs to be improved.

Just now several International experiments are under preparation. They are Double Chooz [10] in France, Daya Bay [11] in China and RENO [12] in Korea. In all 
of them the detectors of a new generation will be used. These detectors can give a possibility of obtaining high statistics while measuring antineutrino spectrum. 
And the high statistics admits to isolate individual fissile isotopes spectra and compare them with measured by conversion technique. The appearing estimated uncertainty 
could be close to the uncertainty of ILL spectra.

We know about some projects with a goal to measure spectra ratio for beta particles of $^{235}$U and $^{239}$Pu [13], which can also help to understand better spectra behavior. 
In the article we consider the method of isotopes spectra isolation by using the direct measurement of positron spectrum from inverse beta decay reaction. As a result we hope 
to obtain $^{235}$U and $^{239}$Pu antineutrino spectra, which produce about 90\% of total antineutrino flux of nuclear reactor.

\section{Antineutrino registration}

Antineutrino can be registered through the inverse beta decay reaction on proton which has the largest cross section
\begin{equation}
\bar{\nu_{e}}+p \rightarrow n + e^{+}.
\end{equation}

The positron appeared as a result of the reaction carries out practically all antineutrino energy [14, 15]. Its kinetic energy is linearly connected with antineutrino energy
\begin{equation}
T=E-\Delta -r_n,
\end{equation}
where $T$ -  positron  kinetic energy, $E$ -  antineutrino energy, $\Delta$ - the reaction threshold equals to 1.806 MeV and $r_n$ is neutron recoil energy.

So, the positron spectrum is the same as antineutrino's, but shifted on 1.8 MeV and convoluted with cross section. Recoil energy in the first approximation can be neglected.

\section{Antineutrino spectrum}

Antineutrino spectrum is being formed inside the reactor core from a number of beta-decays of fission fragments. Fragments are produced by several fissile isotopes, 
not only  $^{235}$U as it was considered at the beginning of the period of searching neutrino. We know that the most part of fissions is produced by four isotopes, 
they are  $^{235}$U,  $^{239}$Pu,  $^{241}$Pu and $^{238}$U.

\begin{figure}
\includegraphics{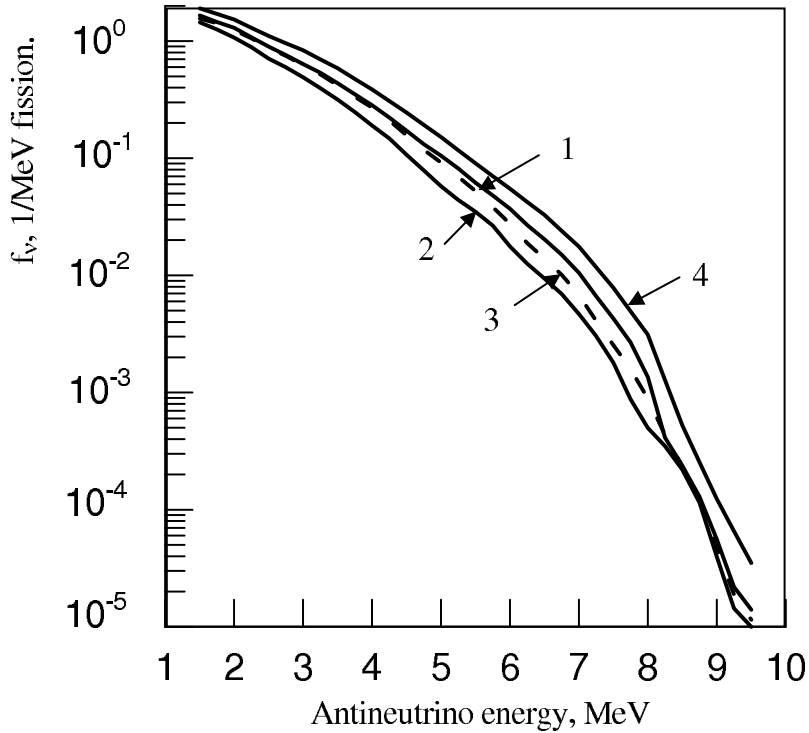}
\caption{\label{fig:epsart} Antineutrino spectra $^{235}$U (1), $^{239}$Pu (2), $^{241}$Pu (3) from [7-9] and $^{238}$U (4) from [18].}
\end{figure}

Each isotope antineutrino spectrum can be calculated using data base for fissile fragments
\begin{equation}
\rho_{\nu}(E)=\sum_i{y_{i}A_{i}(E)},
\end{equation}
where $y_i$ - yield of $i$-th fragment in fission and $A(E)$ is its antineutrino spectrum. But the accuracy of evaluation the spectrum (3) is limited by the number of nuclei 
with unknown decay schemes (about 25\% of total number). They have half-life periods less then 0.3 seconds.

One can take antineutrino spectrum also by another method, by converting experimentally measured beta-spectrum from fissile isotope. This method is accounted 
for the present moment as the most exact. It was used in experiments in ILL (Grenoble). In [7, 8, 9] the beta-spectrum was measured with high level of accuracy for 
three isotopes $^{235}$U,  $^{239}$Pu and $^{241}$Pu , which undergo fission through absorption of thermal neutrons. Fission of $^{238}$U goes on only by 
capturing fast neutrons and has a small fission cross section. Its high enough rate of fissions is explained by the great mass of this isotope in the content of nuclear fuel. 
They fit the measured beta spectra by a set of 30 hypothetical beta-spectra with boarding energies uniformly distributed from 0 up to high energy of experimental spectrum. 
The coefficients found (similar to $y_i$ in (3)) were used in formula (3) for calculating the total antineutrino spectrum.

These antineutrino spectra are accounted as a standard for the moment for data analysis in experiments with reactor antineutrinos. They are presented at figure 1.

There is also the third method of taking antineutrino spectrum, it is a method of direct measuring. When one measures spectrum of positrons from reaction (1) and 
extracts antineutrino spectrum. This method was realized in experiments of Rovno group [16, 17]. The authors have obtained the antineutrino spectrum as an exponential 
function with polynomial of 10-th power while solving the equation
\begin{equation}
S_e(T)=\int{\rho_{\nu}(E){\sigma}_{{\nu}p}(E)R(E,T)dE},
\end{equation}
where $S_e(T)$ $-$ positron spectrum from reaction (1), $\rho_{\nu}(E)$ $-$ antineutrino spectrum, $\sigma_{{\nu}p}(E)$ $-$ inverse beta decay reaction cross section and 
$R(E,T)$ $-$ response function of detector. As a result we have a formula for antineutrino spectrum
\begin{equation}
\rho_{\nu}(E)=5.09\cdot exp(-0.648E-0.0273E^2-1.411(E/8)^{10}),
\end{equation}
This spectrum corresponds to some reactor fuel compositions like the following one
\begin{equation}
^{235}{\rm U} - 0.586, ^{239}{\rm Pu}- 0.292, ^{238}{\rm U} - 0.075, ^{241}{\rm Pu} - 0.047.
\end{equation}

Response function was simulated by Monte Carlo method. This function transforms positron energy spectrum appeared in (1) in experimentally observed one. Each 
detector has its individual response function depending on detector features. At figure 2 one can see response functions for some values of positron kinetic energy for 
detector RONS used at Rovno experiment.

The control for simulated function was done by comparing the spectra measured and calculated from some gamma sources ($^{60}$Co and $^{24}$Na) which were 
placed in the center and the periphery of the detector. Calibration of the detector was made by the beta-source $^{144}$Ce$-^{144}$Pr with boarding energy 2997 keV.

\section{The extraction of $^{235}$U and $^{239}$Pu antineutrino spectra from measured positron spectrum}

During the reactor operational run antineutrino spectrum changes its shape. At the beginning of run the spectrum is formed mainly by $^{235}$U ($\sim$60-70\%) and at the end 
of run the largest or equal with $^{235}$U part of fissions comes from $^{239}$Pu ($\sim$40-50\%). This is a foundation for proposed method of extraction the partial spectra.

Let's regard the positron spectrum (1), which could be measured in a detector placed in the vicinity of some nuclear reactor. At figure 3 one can see positron spectra produced by 
pure isotopes of uranium and plutonium and real spectrum corresponding to (6), which one can observe during reactor operational run. This real spectrum takes place between 
spectra of $^{235}$U and $^{241}$Pu. But at the beginning of the run it will be closer to $^{235}$U and at the end closer to $^{239}$Pu..

\begin{figure}
\includegraphics{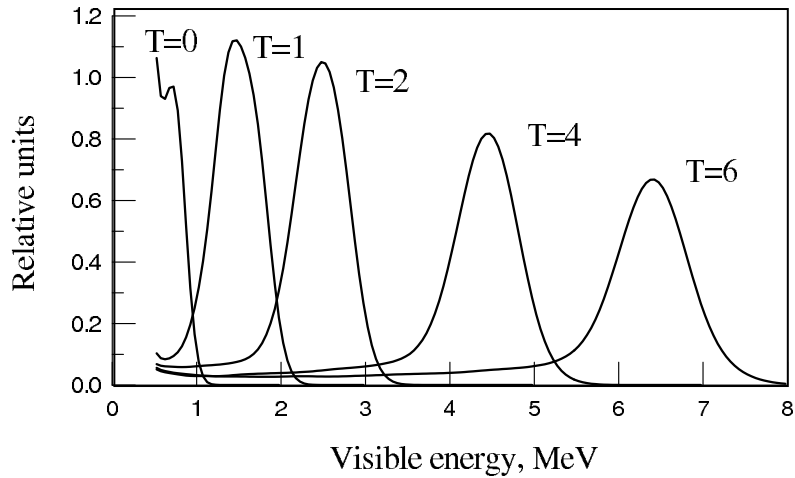}
\caption{\label{fig:epsart} Simulated response function for detector RONS, which was used in Rovno experiments.}
\end{figure}

So, one can divide all the data measured during the reactor run into two parts - ``beginning'' and ``ending''. And we can use them for extracting individual spectra.

Spectrum ``beginning'' contains $^{235}$U in larger proportion. Let's account the input of other fissile isotopes as a background and remove it.

The main question will concern the value of uncertainty of the spectrum. We can suppose that the total uncertainty may be about 1\% including 0.1-0.2\% statistics, as 
they account to achieve in modern experiments like Double Chooz. Expected statistics may be about 106 events. In this case the main error will come from the background where 
the greatest part will appear from the spectra of other fissile isotopes.

We can write the total uncertainty of three "background" spectra like
\begin{equation}
\sigma_b=\sum_{i=8,9,1}{\alpha_{i}\sigma_i}.
\end{equation}

While experimental spectrum is known with high accuracy ${\sigma}_{e}\sim$1\%, for extracted $^{235}$U spectrum we will get the error
\begin{equation}
\sigma_5=\frac{1}{\alpha_5}\sqrt{\sigma_e^2+\sigma_b^2}.
\end{equation}

In (7) and (8) the letter $\alpha_i$ is assigned for individual parts of fission. One can find values of $\alpha_i$, which we have used for the estimations of uncertainties, in table 1.

\begin{table}[h]
\caption{Parts of isotope fission at the beginning and end of reactor operational run}
\label{table:1}
\vspace{10pt}
\begin{tabular}{l|c|c|c|c}
\hline
Isotope  & $^{235}$U & $^{239}$Pu & $^{238}$U & $^{241}$Pu \\
\hline
``beginning'' & 0.65 & 0.25 & 0.07 & 0.03 \\
``ending''  & 0.35 & 0.50 & 0.08 & 0.04 \\
\hline
\end{tabular}\\[2pt]
\end{table} 

It is important to note that the experimental error is not a constant for a whole spectrum, it varies from bin to bin. At figure 4 one can see the standard behavior of experimental 
uncertainty for restored spectrum. At minimum it is equal to 1\% as we supposed to get.

In table 2 we show achievable values of uncertainties for$^{235}$U. One can compare these errors with uncertainties for $^{235}$U spectrum from ILL shown in the second 
column. To obtain the standard spectrum we can have an average of the both spectra and this is shown in the fourth column.

\begin{table}[t]
\caption{Expected relative uncertainty for $^{235}$U antineutrino spectrum (fission part 65\%, see table 1)}
\label{table:2}
\vspace{10pt}
\begin{tabular}{l|c|c|c}
\hline
$E$m MeV & ${\delta}S_5$ (ILL 68\% CL) & ${\delta}S_5(experim)$ (68\% CL) & ${\delta}S_5(average)$ (68\% CL) \\
\hline
2.0 & 0.026 & 0.923 & 0.026 \\
2.5 & 0.024 & 0.155 & 0.024 \\
3.0 & 0.023 & 0.035 & 0.019 \\
3.5 & 0.021 & 0.025 & 0.016 \\
4.0 & 0.020 & 0.027 & 0.016 \\
4.5 & 0.020 & 0.028 & 0.017 \\
5.0 & 0.024 & 0.028 & 0.018 \\
5.5 & 0.026 & 0.029 & 0.019 \\
6.0 & 0.030 & 0.029 & 0.021 \\
6.5 & 0.035 & 0.031 & 0.023 \\
7.0 & 0.040 & 0.055 & 0.032 \\
7.5 & 0.047 & 0.109 & 0.040 \\
8.0 & 0.061 & 0.138 & 0.056 \\
8.5 & 0.134 & 0.276 & 0.120 \\
9.0 & 0.486 & 0.843 & 0.421 \\
9.5 & 0.608 & 1.610 & 0.569 \\
\hline
\end{tabular}\\[2pt]
\end{table} 

Similar table 3 is constructed for $^{239}$Pu, where we used uncertainty from the last column of table 2, while applying the same procedure for this isotope for experimental 
spectrum marked ``ending''.

\begin{table}[t]
\caption{Expected relative uncertainty for $^{239}$Pu antineutrino spectrum (fission part 50\%, see table 1)}
\label{table:3}
\vspace{10pt}
\begin{tabular}{l|c|c|c}
\hline
$E$m MeV & ${\delta}S_9$ (ILL 68\% CL) & ${\delta}S_9(experim)$ (68\% CL) & ${\delta}S_9(average)$ (68\% CL) \\
\hline
2.0 & 0.027 & 1.200 & 0.027 \\
2.5 & 0.026 & 0.204 & 0.026 \\
3.0 & 0.026 & 0.055 & 0.024 \\
3.5 & 0.026 & 0.048 & 0.023 \\
4.0 & 0.027 & 0.057 & 0.024 \\
4.5 & 0.029 & 0.064 & 0.026 \\
5.0 & 0.032 & 0.074 & 0.029 \\
5.5 & 0.036 & 0.074 & 0.032 \\
6.0 & 0.041 & 0.090 & 0.038 \\
6.5 & 0.045 & 0.100 & 0.041 \\
7.0 & 0.067 & 0.183 & 0.063 \\
7.5 & 0.116 & 0.229 & 0.103 \\
8.0 & 0.213 & 0.438 & 0.191 \\
8.5 & 0.486 & 0.266 & 0.234 \\
9.0 & 0.578 & 1.320 & 0.529 \\
9.5 & 0.608 & 2.340 & 0.588 \\
\hline
\end{tabular}\\[2pt]
\end{table} 

\begin{figure}
\includegraphics{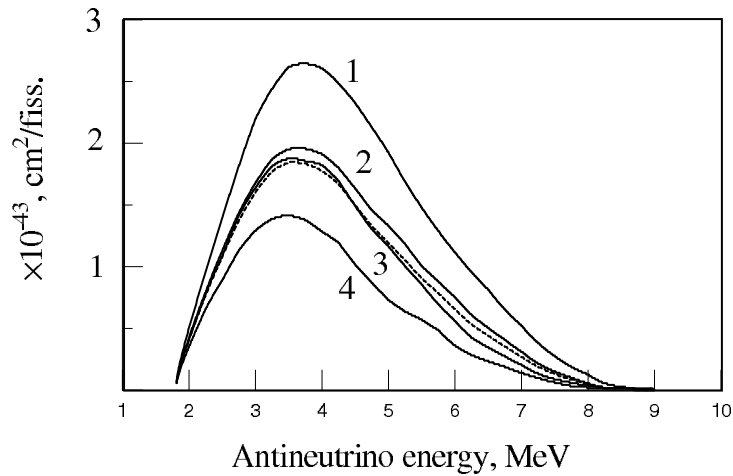}
\caption{\label{fig:epsart} Positron spectra: 1 -  $^{238}$U, 2 - $^{235}$U, 3 - $^{241}$Pu, 4 - $^{239}$Pu. Dashed line shows spectrum corresponding to 
the fuel composition (6).}
\end{figure}

The last table 4 demonstrates what the values can be achieved while applying the procedure to $^{238}$U. This spectrum we can get to know only from calculations. 
There may be a good chance to take it experimentally.

\begin{table}[t]
\caption{Expected relative uncertainty for $^{238}$U antineutrino spectrum}
\label{table:4}
\vspace{10pt}
\begin{tabular}{l|c|c|c}
\hline
$E$m MeV & ${\delta}S_8$ (Vogel 68\% CL) & ${\delta}S_8(experim)$ (68\% CL) & ${\delta}S_8(average)$ (68\% CL) \\
\hline
2.0 & 0.05 & $-$ & 0.05 \\
2.5 & 0.06 & 1.270 & 0.06 \\
3.0 & 0.06 & 0.301 & 0.06 \\
3.5 & 0.08 & 0.192 & 0.074 \\
4.0 & 0.10 & 0.188 & 0.088 \\
4.5 & 0.10 & 0.187 & 0.088 \\
5.0 & 0.10 & 0.186 & 0.088 \\
5.5 & 0.10 & 0.195 & 0.089 \\
6.0 & 0.10 & 0.197 & 0.089 \\
6.5 & 0.10 & 0.195 & 0.089 \\
7.0 & 0.20 & 0.294 & 0.165 \\
7.5 & 0.20 & 0.508 & 0.186 \\
8.0 & 0.30 & 0.778 & 0.280 \\
8.5 & 0.40 & 0.914 & 0.366 \\
9.0 & 0.70 & $-$ & 0.696 \\
9.5 & 1.00 & $-$ & 0.997 \\
\hline
\end{tabular}\\[2pt]
\end{table} 

\begin{figure}
\includegraphics{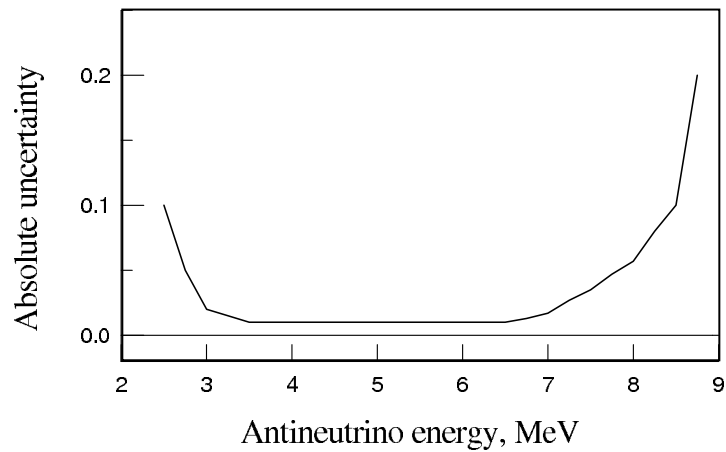}
\caption{\label{fig:epsart} Experimental uncertainty of antineutrino spectrum after transforming from positron spectrum. Suppose that systematic and statistical 
errors in total do not exceed 1\% in the most part of spectrum.}
\end{figure}

\section{Conclusion}

The discussed method of obtaining $^{235}$U and $^{239}$Pu antineutrino spectra can improve uncertainties in their spectra. It is important for neutrino control method. 
The using of different methods of spectra obtaining increases the reliability of standard spectra. The accuracy in 3\% in not enough today for measuring the fuel composition 
by neutrino method, 1\% seems desirable. In that case statistics 2000$-$3000 events/day may be enough to have uncertainty ~5\% for $^{235}$U part of fission per month 
of measurement.

From one hand this method is additional to other methods of spectra obtaining, from the other hand it is direct and is not affected by some procedures of recalculating like 
converting method. Also it does not depend on knowledge of decay schemes as calculating method because it accounts all decays automatically.

In collaboration with ILL spectra there may be achieved better accuracy for standard spectra. Also there may be directly determined the $^{238}$U spectrum in 
spite of the high uncertainty and compared to the calculated one.

Of cause the direct method is additional to other methods. These calculations demonstrate the importance of continuation of the experiments on measuring beta-spectra 
of the fissile isotopes and improving the conversation procedure because it is the most exact. Direct measurement may serve as an indirect test for conversion spectra. 
Calculations are very important for searching time evolution of nuclear reactor antineutrino spectrum. All three methods improve the complete knowledge of reactor 
antineutrino spectrum and its behavior during the operational run. Altogether, the usage of several methods increases reliability of partial spectra while proposing the standard 
spectra for the future analysis. As an example we can attract the attention to hard part of ILL spectra (higher than 8 MeV) where all spectra become the same, while calculated 
spectra demonstrate growing difference in this region.

Proposed method on base of Rovno experience shows its applicability in the analysis of the data in future Double Chooz experiment.

\section*{Acknowledgments }

Author thanks L.A. Mikaelyan and A.Ya. Balysh for useful discussions and friendly criticism.

\end{document}